\documentclass{article}

\usepackage{subfigure}

\usepackage{natbib}
\bibpunct[, ]{(}{)}{;}{a}{}{,}
\renewcommand\bibfont{\fontsize{10}{12}\selectfont}

\usepackage{graphicx}

\usepackage{shortcuts}
\usepackage{tabularx}
\usepackage{shortcuts}
\usepackage{makecell}
\usepackage{xcolor,colortbl}
\usepackage{hyperref}
\definecolor{Gray}{gray}{0.85}
\definecolor{LightCyan}{rgb}{0.88,1,1}

\newcommand{\mc}[2]{\multicolumn{#1}{c}{#2}}

\definecolor{ColOne}{rgb}{0.8571428571428571, 0.8571428571428571, 1.0}
\definecolor{ColTwo}{rgb}{1.0, 0.8571428571428572, 0.8571428571428572}

\newcolumntype{a}{>{\columncolor{ColOne}}r}
\newcolumntype{b}{>{\columncolor{ColTwo}}r}

\newcolumntype{Y}{>{\centering\arraybackslash}X}

\begin{document}

\date{}
\author{
Sabina Tomkins\\
  \texttt{stomkins@umich.edu}
  \and
David Rothschild\\
  \texttt{david@researchdmr.com}
  \and
  Alex Liu\\
  \texttt{avliu@umich.edu}
  \and
  Alexander Thompson\\
  \texttt{alexmt@umich.edu}
  }

\title{Identity isn't everything - how far do demographics take us towards self-identified party ID?}

\maketitle

\begin{abstract}
How well do demographics explain party identification? 
Demographics are related to party identification
in  
political polls, news articles, and  academic publications. 
Yet, there is a diversity of party identification even within demographic groups which have historically been attached to one party. 
And some groups lack a clear connection to either party.  
It may be
that demographics on their own fail to account for the fact 
that people generally belong to a variety of groups. 
They must select the groups which are most important to them when 
shaping a political identity, and may choose to construct an identity 
relatively unattached to any specific demographic group to which they belong. 
This prompts the question, 
do we need to consider measures of identity strength when 
using demographics to explain party identification?
We utilize a predictive framework to address these questions 
and find that demographics are highly predictive for some groups (e.g., Black Democrats), 
while others benefit from the inclusion of identity strength (e.g., Hispanic Republicans). 
\end{abstract}

\section{Introduction}

Partisanship has increasingly become a salient feature of social group identity \citep{mason2015disrespectfully}, with partisans increasingly internalizing party identification into their sense of self \citep{west2022partisanship,egan2020identity,margolis2018politics}. Consequently, party identification can drive policy positions \citep{jenke2020voter} (rather than the other way around), influence vote choice even in nonpartisan elections \citep{bonneau2015party}, and drive other, non-political, behavior \citep{manento2022party,rau2022partisanship,huberman2018economic,huddy2017political}. The rise in party affiliation as an aspect of identity has also been attributed to polarization \citep{iyengar2012affect}. That is, 
in-group partisans view members of the other party (out-group members) with increasing bias and negative affect \citep{iyengar2012affect}. Thus, 
determining and understanding 
 party identification as a function of social identity has important practical and scientific insights, with implications from marketing to broad political phenomena, such as threats to democracy. 

Yet, any attempt to relate social group identities to partisanship is impeded by competing measures of social group identity (i.e., what measures should a researcher consider), along with a lack of comprehensive work which investigates interactions between these measures across demographics \citep{conover1988role,lee2008race}.
Demographic factors (such as age, gender, education and 
income)
have been show to be indicators of party identification (party id) in a variety of contexts 
and over time
\citep{knoke1974social,trevor1999political,green2004partisan,huerta2021red,berry2021white}.
However, these analyses and the relationship between demographics and party id 
more broadly suffer from the fact that belonging to a group is only one facet of political identity. 
In addition to group belonging, 
Perceived Membership and Linked Fate
have been argued as constituting important and orthogonal dimensions of political identity 
\citep{mcclain2009group}.

Our goal is to understand the extent to which demographics can predict party id. 
We find that for some groups (groups with homogeneous support for a party), simple demographic information is sufficient to predict party id. 
For others (relatively more heterogeneous groups), it is woefully insufficient. 
For these groups, 
we utilize a range of identity strength measures (which we term \textsc{Identity+} measures). 
We show that including \textsc{Identity+} measures can improve the predictions for minorities within a party by upwards of 20\% on average.

\section{Predicting Party Id with Demographics}
We begin by predicting party id for all voters in the 2020 American National Election Studies \citep{anes}. 
To do so, we introduce the nomenclature \textsc{Standard Demographics}. These describe memberships in groups which are typically found to be related to party id: racial groups, age groups, gender groups, and education groups. 
For each respondent in the ANES data, we form a dataset with their race, age, gender, and education, and party id which we label as one of: Democrat, Independent, and Republican. 

 \tabref{fig:ANES_Prodege} shows the confusion matrix our data in predicting Democrats, Republicans, and independents with \textsc{Standard Demographics} (defined above). On the ANES data, we obtain a recall of 0.63 and precision of 0.53 for Republicans. 
On ANES we obtain a  recall 0.63 and precision of 0.59 for Democrats. The F1 score is the harmonic mean of precision and recall: 2*precision*recall/(precision + recall). The F1 in ANES is 0.58 for Republicans and 0.61 for Democrats. 
Henceforth, we use F1 scores as the main metric of interest.

\begin{figure*}[htp]
\centering
\includegraphics[width=\linewidth]{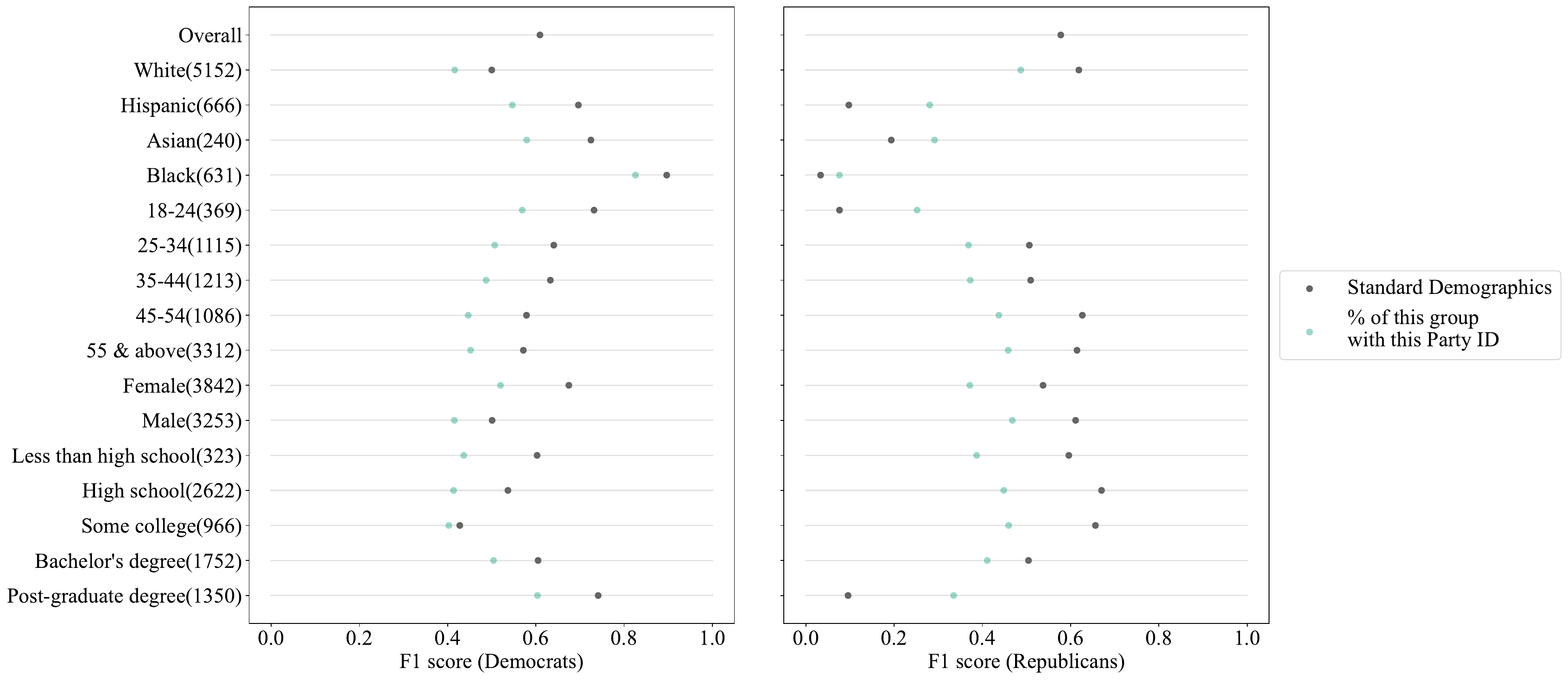}
\caption[F1-score by group for all ANES]{The F1 score on different sub-groups for both Democrats and Republicans. The number of respondents in each group is shown in parentheses. We see that the gains by race vary greatly depending on the party alignment of the majority of members of that race group. Models were run on a total of 7887 respondents. We also show for each group, the proportion of members of that group who identify with each party. 
}
\label{fig:1WayANES}
\end{figure*}

Next, we turn to differences between groups. In \figref{fig:1WayANES}, 
we see that belonging to some groups is more predictive than belonging to others. 
For example, Black democrats are easily identifiable. 
To the contrary, the F1 score for Black republicans is very low.

However, although some groups can be easily assigned a party id, overall, the 
F1 scores of .61 for democrats and .58 for republicans are fairly low. 
These simple demographics  only predict party id slightly better 
than random chance, and for some groups they are fairly useless. 
This prompts the question, can incorporating identity strength 
 better relate demographics to party id?

\section{Demographics, Identity Strength, and Party Id}
Group belonging or membership is only one dimension 
of political identity. 
We may fare better at predicting party id if 
we develop a more personalized model of identity. 
Scholarship on the relationship between social group identity and  partisanship \citep{campbell1980american} is extensive. This work stresses not only the role of individual identities, such as race, to inform partisanship, but the integrative role of partisanship in collapsing multiple identities into a party affiliation \citep{mason2018one}. However, the role of choice in identity \citep{huddy2001social} can complicate this integration when individuals are influenced by a range of potentially conflicting social group identities \citep{brader2014cross,bergersen2018intersectionality,smith2007social,matos2023politics,bedolla2007intersections}. 

Considering social group identities in isolation, race has been a consistently dominant feature of American politics \citep{dawson1995behind,hutchings2004centrality,perez2021diversity}. One's race and racial attitudes can shape their policy preferences \citep{craig2018racial,krupnikov2016political,jardina2022polls}, and political behavior \citep{knuckey2015racial}.  It is linked to partisanship in profound and historically complex ways \citep{acharya2018deep}; for example, 
race and partisanship are currently so firmly linked in the American subconscious that activating associations for an out-race associated with an out-party can increase negative affect for the out-party without it ever being explicitly activated \citep{westwood2020inseparability}.

While the importance of race to political identity is rarely understated, its measurement is contested. A common starting point is \textbf{\textit{Group Membership}} (i.e., what group, or groups, a person assigns themselves to within all demographic categories). In explaining Black politics in the United States, Dawson originated the term \textit{linked fate} \citep{dawson1995behind}. Linked fate describes the extent to which group members see their fate as linked to the fate of the group. He argues that even as their economic conditions change, 
Black voters see their fate linked to the fate of Black people in the United States. Linked fate, which we call \textbf{\textit{Group Fate}} has remained an essential measure of racial identity in politics \citep{simien2005race}. \citep{mcclain2009group} added a third measure: perceived membership, which we term \textbf{\textit{Group Association}} (i.e., which groups a person associates themselves with). 

These models, largely developed for Black voters, have been questioned for Hispanic and Asian voters \citep{junn2008asian,gershon2019intersectional}. However, that is not to say that race does not factor into politics for these groups  \citep{sadhwani2022influence,cox2019most}. 
For Asian and Hispanic voters the groups which they choose to ignore may be just as significant as those they choose to incorporate into their identities \citep{basler2008white,junn2008asian}. Recently, Hickel et al. found that Hispanic voters who identified as American, rather than Hispanic, were more likely to vote Republican \citep{hickel2020role}.

Simultaneously, scholars have called for a new understanding of white political identity \citep{jardina2019white}. Unlike previous work which has focused on out-group animus, Jardina has proposed that white identity is also explained by in-group attitudes, that is, those with higher white identity value being white, and have positive attitudes towards other whites. 
Additionally, recent work has investigated the extent to which some whites may see their group as a dispossessed minority \citep{jardina2021group}, suggesting the importance of group consciousness for this group as well \citep{berry2019white}.

The rise of the narrative of oppression in  political communication, along with the associated supports for populism in recent elections, 
bring up questions around the extent to which perceived societal standing 
influences political identity and behavior \citep{mondon2019whiteness,berlet2019rural,kazin2016trump}. 
Gest et al. measured subjective social, political, and economic status to be able to predict support for the Radical Right in the United States and United Kingdom \citep{gest2018roots}.  They found that nostalgic deprivation, a construct tightly related to ideas of a dispossessed minority, predicts support for the Radical Right. Whereas measures of societal standing are not generally included in models of partisanship, we incorporate this in a measure we term \textit{\textbf{Group Standing}}.

\section{Empirical Strategy}
To understand the advantages of incorporating identity strength into demographics based models 
of party id, we design a survey around a range of identity measures:

\begin{itemize}
    \item \textbf{Group Membership} We start the survey by asking participants to select subgroup membership within 12 categories (and party identification):  age, gender, race and ethnicity, education, income,  religion, LGBTQ status, immigration status, union status, veteran status, place of residence, industry, and political party. 
    \item \textbf{Group Association}
To determine which groups the respondent most identifies with, we ask, ``These descriptions matter the MOST to me'' and ``These descriptions matter the LEAST to me''. The question is structured such that a respondent must select three of the possible subgroups that matter the most and three which matter the least. 
Each of the possible responses are generated  given the respondent's previous answers.  For example, a woman who wrote in her race as Salvadoran,\footnote{For this and the other identity strength items, if respondents indicated belonging to multiple groups we asked them to indicate which one they prefer that we refer to for the rest of the survey. So if someone indicated that they were both 'Black' and 'Vietnamese' we would ask them to indicate which group they prefer we use for the rest of the survey to refer to their race. We view this as a modest improvement over options which either force people to select one race, or refer to race in abstract terms. However, in future work we will refine the delivery of the survey to allow people to better describe multiple memberships. } with a college education, who works in the health industry, is a practicing protestant, and a union member, could say that the three most or least important descriptions to her are: ``I am Salvadoran'', ``I am a union cardholder'' and  ``My college degree''. This departs from surveys which allow respondents to select groups without identifying subgroups, for example, by selecting race/ethnicity, union status, or education as being important to them.

\item \textbf{Group Fate} Here, we follow a similar approach as for group association. For each group that a participant belongs to they must choose one of the following: “What happens generally to THIS GROUP in this country will have something to do with what happens in my life” and “What happens to this group will NOT impact my life; What happens to this group will impact my life A LITTLE; What happens to this group will impact my life SOMEWHAT; What happens to this group will impact my life A LOT”. We also constrain responses such that for at least one group and no more than three, they must select the most extreme options (NOT and A LOT). 

\item \textbf{Group Standing} To determine how one sees the groups within a societal context, we ask respondents to ``Place each group into categories indicating how important you think they are in American society.'' They are shown a diagram of concentric circles, and told that `` ‘1’ represents those that are considered the most central and important to society, whereas ‘4’ represents those that are considered the least central and important to society.''
This question was adapted from the question developed by Gest et al. \citep{gest2018roots}. In the original conception, people are asked to indicate the relative importance of large demographic groups, such as ``The Elderly''. Here, we are using standing as another dimension of identity, and asking respondents to indicate the relative importance only of those groups which they have indicated belonging to. 

\item \textbf{Group Support}
Finally, we ask participants to indicate how much they would favor different groups they belong to with financial support: both with a donation by a charity and spending by the government.
We ask two questions: ``A charity has allocated funds to target groups of students to send to college and you've been asked to decide who should receive those funds among the following groups. Select the three groups you would target to send to college'' and ``The government has allocated funds to target groups of students to send to college and you've been asked to decide who should receive those funds among the following groups. Select the three groups you would target to send to college:''.
Now, they are shown the same groups as in previous questions, with the groups rephrased to be plural, for example, if a respondent selected \textit{Protestant} earlier, they would be shown the option to give to \textit{People of Protestant Faith}.

\end{itemize}

We group these measures into three different datasets, each of which cumulatively adds variables to the previous set: \textsc{Standard Demographics}, \textsc{Expanded Demographics}, and  \textsc{Identity+}\footnote{Code to repeat our analysis will be shared upon publication.}. 
 These datasets are all formed from the same survey respondents, where we gradually enrich the data with responses to more questions.
 \textsc{Standard Demographics} contains group membership only for the demographics of: race, age, gender, and education. 
 \textsc{Expanded Demographics} contains group membership in the standard demographics, along with income, religion, LGBTQ status, immigration status, union status, veteran status, place of residence,and industry. 
\textsc{ Identity +} contains the responses in expanded demographics (which together comprise Group Membership), along with the additional questions on Group Affiliation, Group Fate, Group Standing and Group Support.  
We reshape group association and fate questions slightly to be relativistic. That is, respondents cannot only express how much a given group matters to them in isolation, they must choose which groups matter when confronted with a set. We believe that this represents a real-life scenario where limited group identities will be activated at a time (e.g. by a particular candidate).

We model partisanship with a seven-point scale and
our final outcome variable is grouped into three buckets: Democrat (Strong Democrat, Weak Democrat, Leans Democrat), Republican (Strong Republican, Weak Republican, Leans Republican), or Independent (Neither).
To mitigate potential bias, this is the final question that respondents see in the section on demographics.

See the full survey document \href{https://drive.google.com/file/d/10GZUlG9EeMFcJu3vM7i8XD-GD2xFmn6w/view?usp=drive_link}{here}.
For modelling purposes we encode race and ethnicity as belonging to a single combined variable with five possible values: white, Black, Asian, Hispanic\footnote{Here, we use the term Hispanic as an aggregate label for respondents who identified as Hispanic, Mexican, Cuban, Puerto Rican, or with  another Hispanic group which they wrote in a free response text box. } or other. However, in the survey respondents are asked about race and ethnicity separately. They are given multiple options for Asian and Hispanic groups, and have an option to further specify Middle Eastern or European ancestry. As less than 1 percent of respondents supplied an ancestry we ignored this field in the analysis.

\subsection{Sample Design}
We conduct a survey with an online panel of adults in the United States\footnote{Deemed exempt by an institutional IRB. Anonymized responses will be shared upon publication. Consent was obtained at the start of the survey and an option to opt-out was provided, more details in appendix}. 
Our data consists of $\sim$2,000 responses obtained with a panel recruited by the firm Protege  utilizing a Qualtrics infrastructure. The sample was collected with quota parameters for age, race, education, gender, and party identification.\footnote{Quotas were: gender (max 1,200 for either gender), age (max 600 for three age groups, but 800 for 45-64), race (max 1,440 for white, max 400 for Black, and max 600 for Hispanic), education (max 800 for high school or less, max 600 for some college, max 800 for college or more), and party max 750 for either party or other (on a 7-point scale).} We obtain a population reasonably similar to the general US adult population ( 70\% of our sample is white, 40\% has some college education but no 4-year degree, 30\% is 65 years old or older, and 40\% is female, compared to 60\%, 30\%, 15\%, and 50\% in the general US population (ACS 2021)). 
These loose quotas do not ensure us a demographically representative response, but we predict on an individual level, limiting the impact of non-representation.

\section{Results}
\begin{figure*}[htp]
\centering
\includegraphics[width=\linewidth]{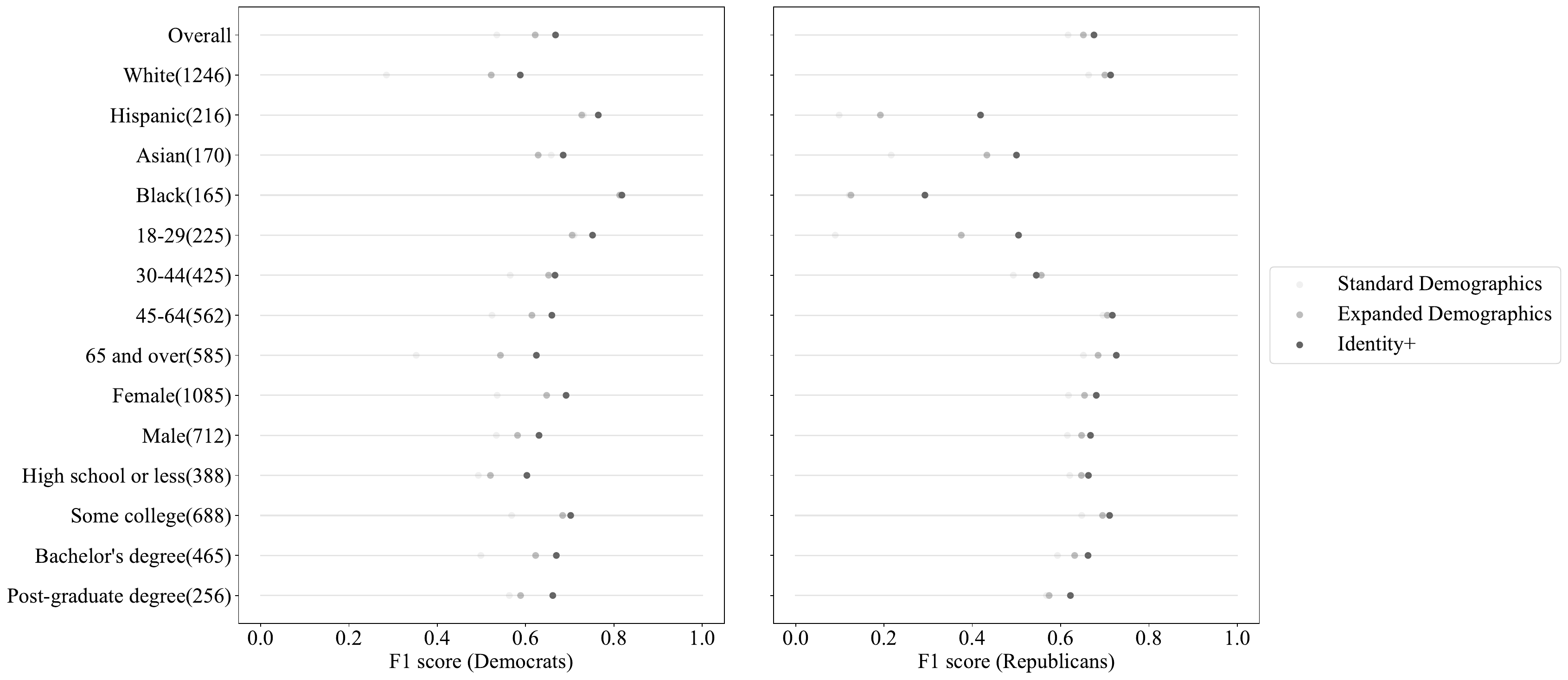}
\caption[F1-score by group for all our data]{The F1 score on different sub-groups for both Democrats and Republicans. The number of respondents in each group is shown in parentheses. We see that the gains by race vary greatly depending on the party alignment of the majority of members of that race group. Models were run on a total of 1997 participants. 
}
\label{fig:1Way}
\end{figure*}

In \figref{fig:1Way} we see that the inclusion of
additional identities and then \textsc{Identity+} factors improves performance overall, with large improvements for groups which are minorities within their party. Overall, we find that with \textsc{Standard Demographics} the overall F1 score for Democrats and Republicans are 0.53 and 0.62 respectively. With \textsc{Expanded Demographics} the F1 scores are .62 and .65, and with \textsc{Identity+} they are 0.67 and 0.67 Democrats and Republicans respectively.  \textsc{Identity+} improves the F1 by 26\% and 9\% over \textsc{Standard Demographics}, for Democrats and Republicans respectively. 

We also see heterogeneity in the advantages to modeling identity strength. 
For example, 
Hispanic, 18-29 year olds, Asian, and Black republicans see significant gains when identity strength is added.
On the democrat side, the gains are most clearly articulated for 
respondents who are white, 65 and older, and who have a high school education or less. 
For some groups, such as  Black democrats and white republicans, there
is basically no advantage to considering identity strength. 

\subsection{Identity Strength Measures}
\begin{figure*}[htp]
\centering
\includegraphics[width=\linewidth]{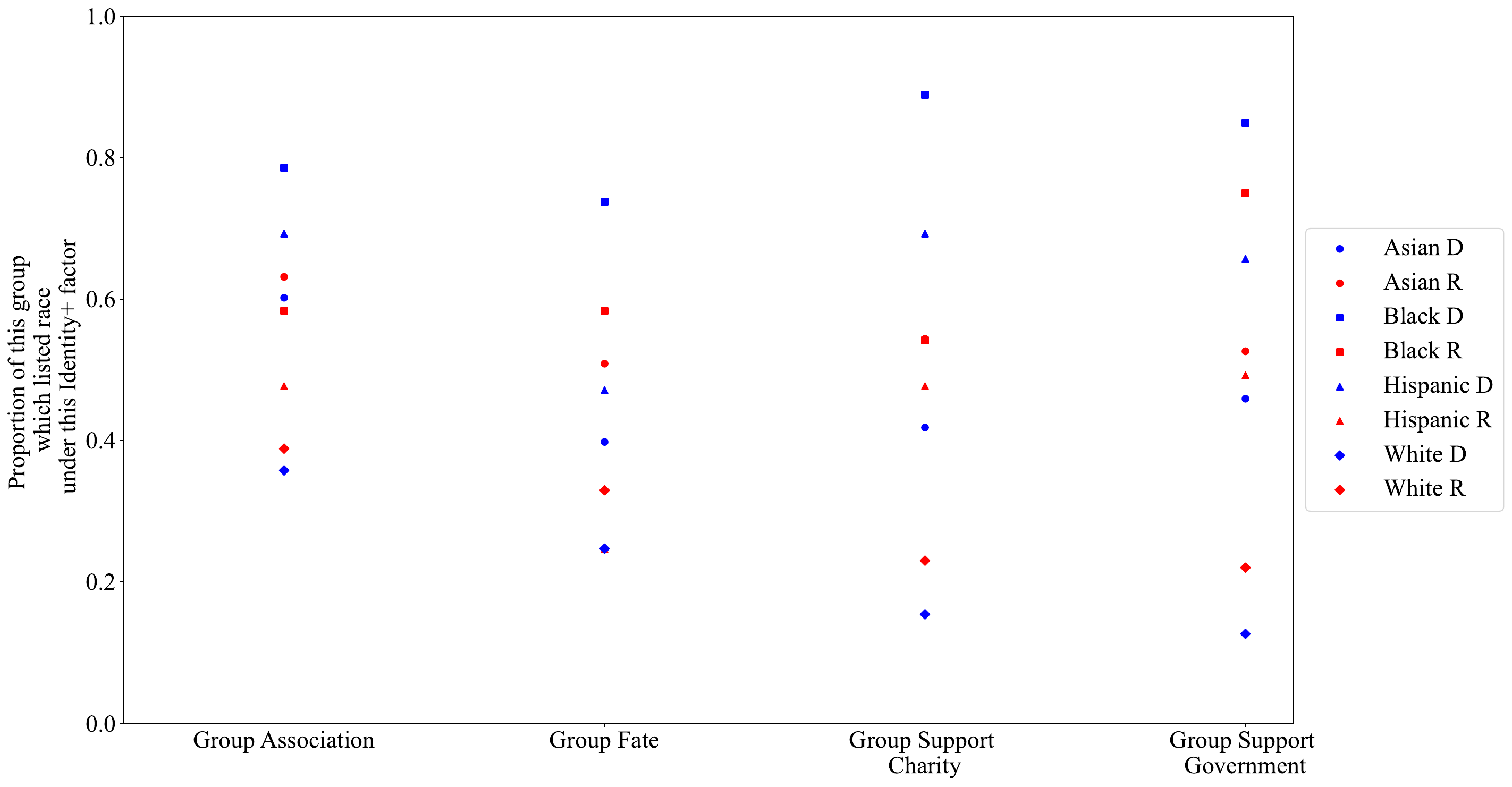}
\caption{Identity strength measures for each race and party id.}
\label{fig:races}
\end{figure*}

\begin{figure*}[htp]
\centering
\includegraphics[width=\linewidth]{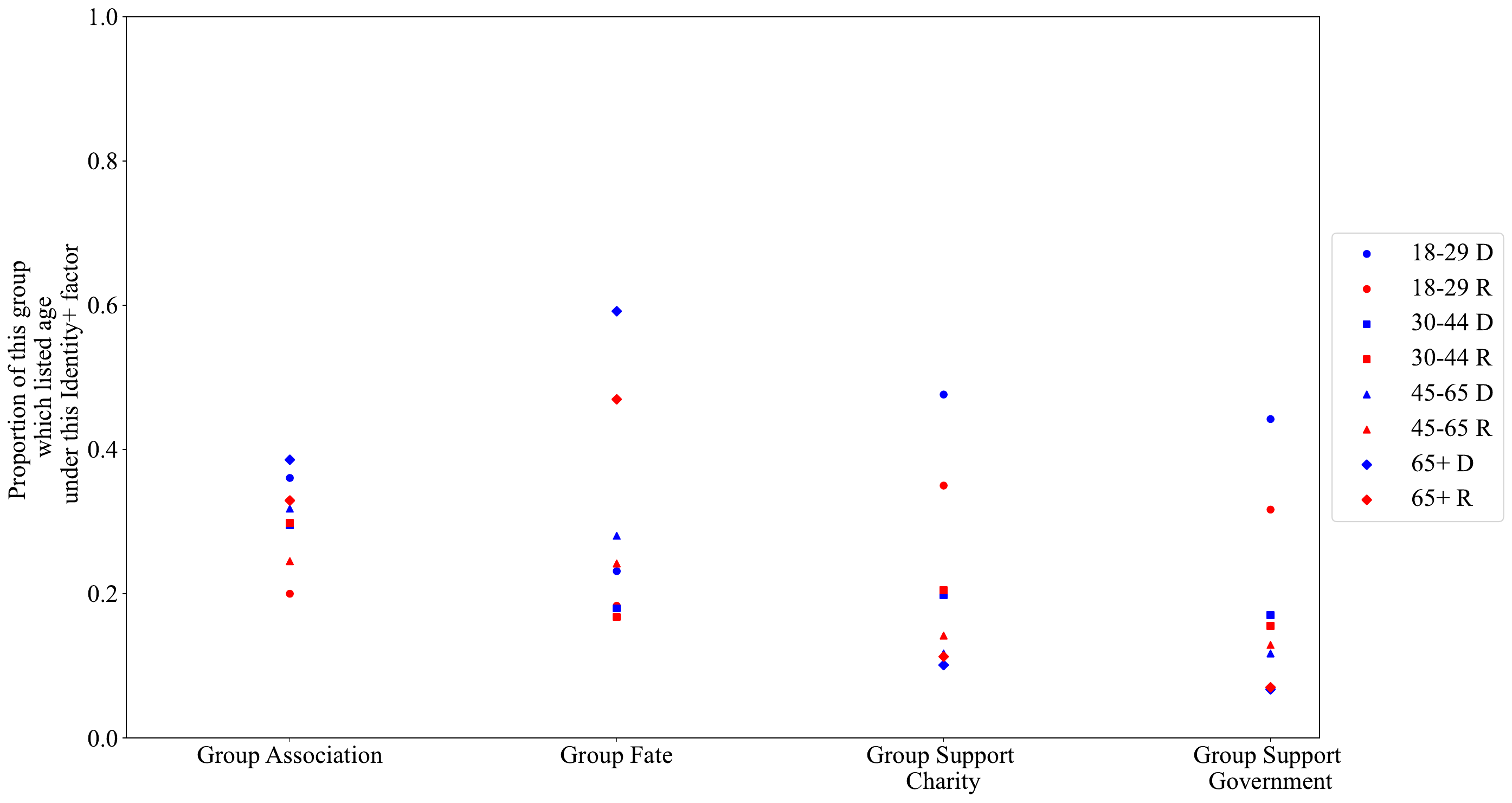}
\caption{Identity strength measures for each age and party id.}
\label{fig:ages}
\end{figure*}

\begin{figure*}[htp]
\centering
\includegraphics[width=\linewidth]{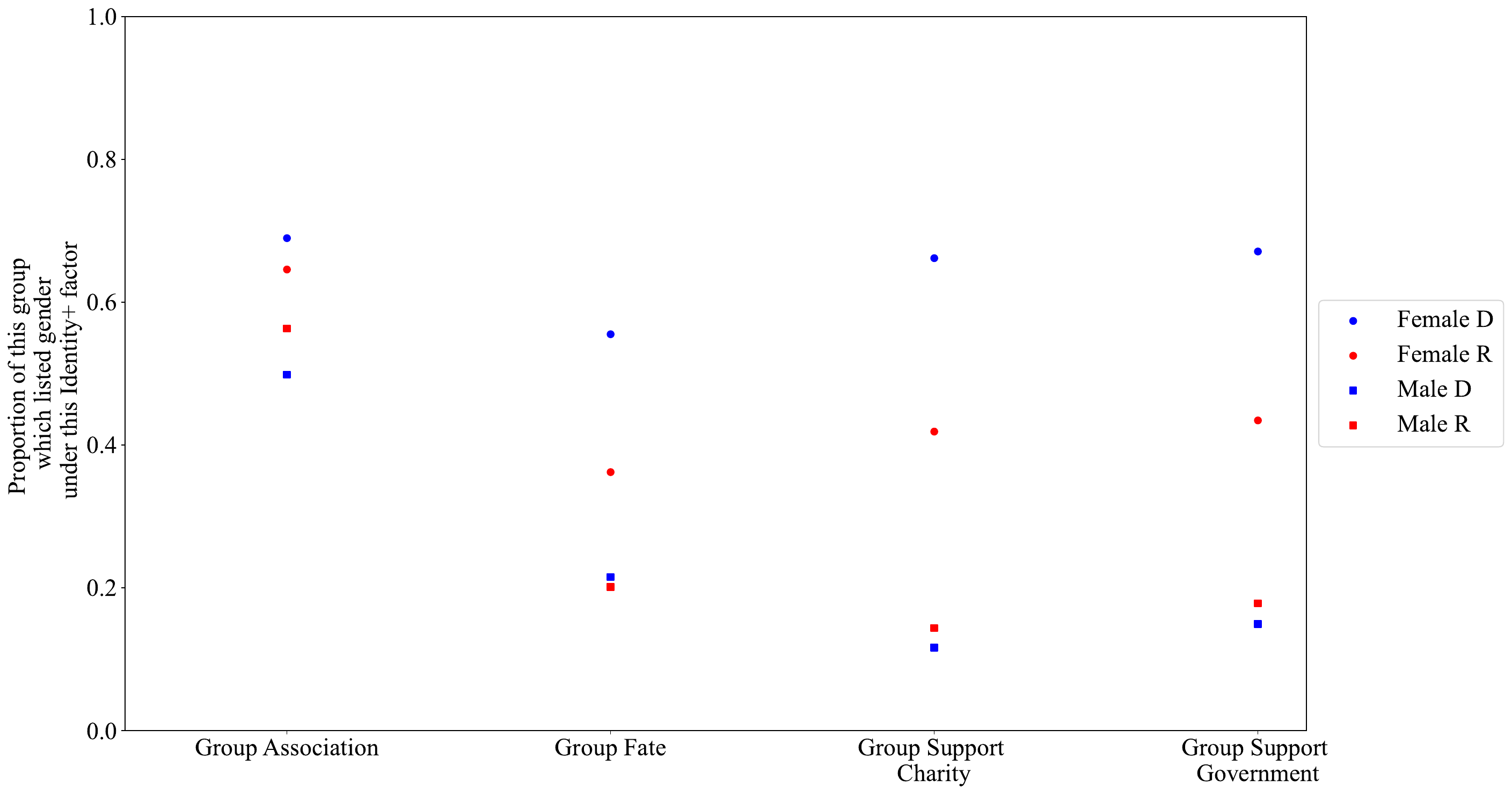}
\caption{Identity strength measures for each gender and party id.}
\label{fig:gender}
\end{figure*}

\begin{figure*}[htp]
\centering
\includegraphics[width=\linewidth]{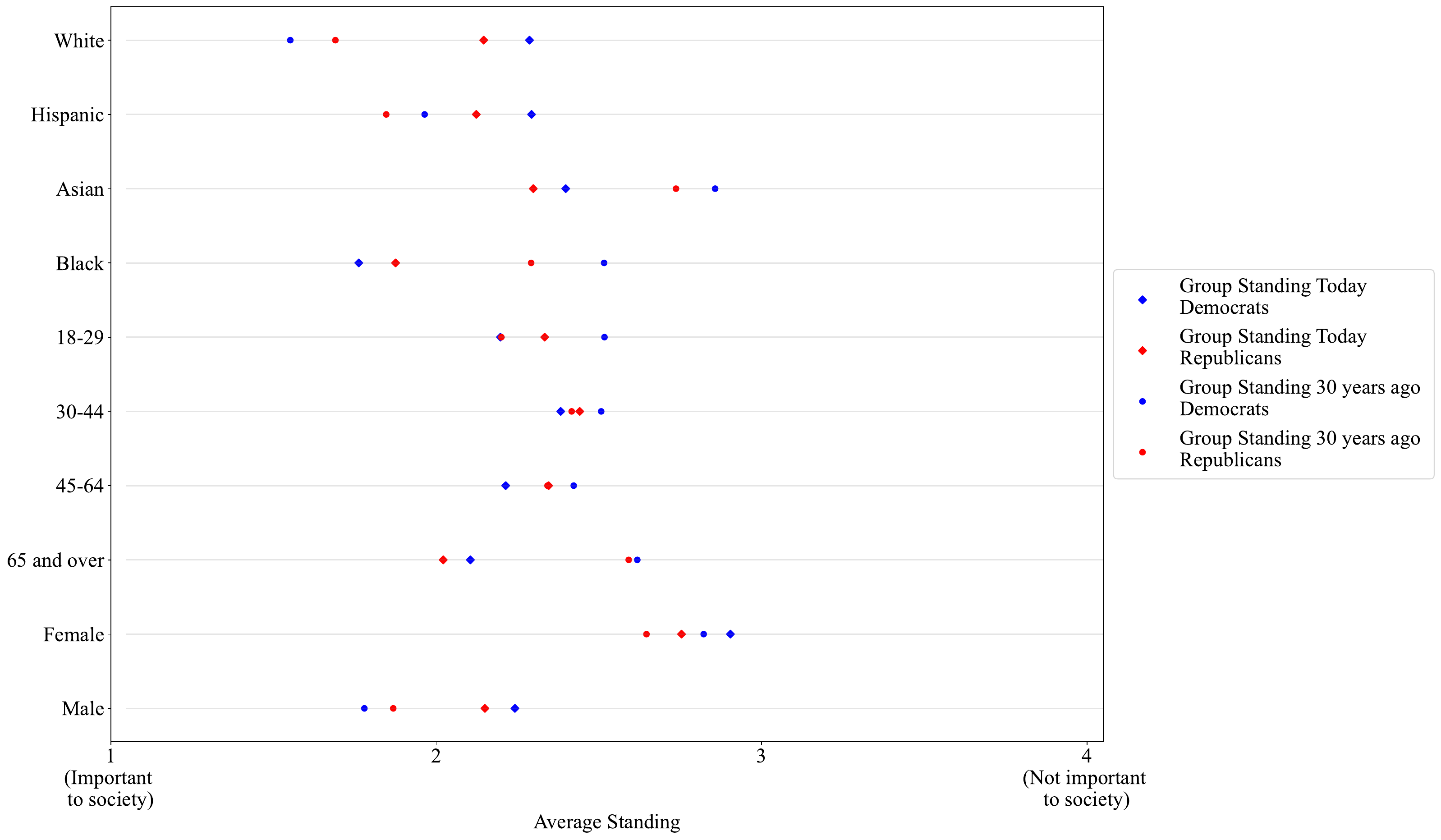}
\caption{Group standing for different groups.}
\label{fig:standing}
\end{figure*}

Why  is identity strength helpful for these groups?
Which identity measures matter?
In \figref{fig:1Way}, 
we see that the  largest performance gains when 
\textsc{Identity+} measures are introduced are enjoyed by certain 
race, age, and gender groups. 
To understand why, 
in \figref{fig:races}, \figref{fig:ages},  \figref{fig:gender}, and \figref{fig:standing} we explore how different groups incorporate
different identity measures. 

In \figref{fig:races} all identity strength measures show wide differences between races and between partisans within a race. 
For example, when it comes to Group Association, Black and Hispanic Democrats are much more likely to associate their race with their identity 
than Black and Hispanic Republicans. 
When it comes to Group Fate, Black respondents of both parties stand out, 
reporting higher amounts of Group Fate compared to respondents from other race groups, 
while there is a clear party difference. 
Asian respondents also diverge by party when it comes to Group Fate. 
When it comes to Group Support, Black Democrats consistently indicate that race is important to them. 
White respondents are more likely to incorporate race into their identity through 
Group Association and Group Fate, than through Group Support. Across all measures, 
white Republicans are more likely to incorporate race into their identity than white Democrats. 

We see less diversity in \figref{fig:ages}. 
Here, the youngest and most elderly groups stand out.
For example, there are clear partisan differences for the 18-29 year old age group 
when it comes to Group Support. This age group is more likely to select age 
as a qualifier for support, but younger Democrats are more likely to do so. 
When it comes to Group Fate, older Democrats are more likely to see their fate 
as tied to others with the same age as other respondents and even Republicans 
of the same age group. 

In \figref{fig:gender}
we see that female Democrats incorporate gender into their identity 
through Group Fate and Group Support much more than female Republicans.
When it comes to Group Association on the other hand, they exhibit similar responses. 
Across the all measures gender is a more common identity for females than for males.

Together, these figures demonstrate why \textsc{Identity+} measures 
allow us to better predict partisanship. 
There are clear differences in how people who belong 
to the same race, age, or gender group think about 
Group Association, Group Fate, and Group Support. 
We see less pronounced differences when it comes to Group 
Standing (\figref{fig:standing}). 

\section{Limitations}
Here, we explore groups in isolation while in fact
respondents create identities at the intersection of different groups, 
especially when those groups share similar societal contexts
\citep{bedolla2007intersections,bergersen2018intersectionality,tormos2017intersectional}. 
Future work might use artificial intelligence to allow respondents to freely describe 
themselves. Such an approach may better elicit the ways in which people associate 
with complex groups which encompass different demographic categories and specifications.

There are many potential measures of social identity which were left out of this analysis. 
For example, we do not explore measures of collective narcissism \citep{de2009collective}.
Rather than follow multi-dimensional models of social identity \citep{cameron2004three,leach2008group}, 
we followed multi-dimensional models of political identity \citep{mcclain2009group}. 
For example, we incorporated two common measures from within the political identity research:
 group association and group fate. 
The measures we explore are not comprehensive, but we hope that given 
the existing literature they can be helpful in moving this line of inquiry forward. 

\section{Discussion}
How well do demographics explain party identification?
Given our experiments with two datasets, 
one standard in the political science literature, 
and one which we collected for this task, 
we see that demographics only go so far in explaining party identification. 
For some groups (e.g., Black respondents) simple demographics are sufficient to predict party id. 
For other groups where party id is more diffuse (e.g., White respondents) simple demographics are not predictive on their own.

Measures which capture not only the groups that one belongs to, 
but how those groups are incorporated into one's identity, 
can improve the predictions of party id. 
In particular, 
we see significant differences in how respondents of each party 
think about Group Association, Group Fate, and Group Support. 

Still, even with a  comprehensive list of identity measures
and the addition of less common demographics (e.g. military and union status), 
we see that we can only approach an F1-score of 0.67 on average
when predicting party id. 
If one's goal is to target likely members of a given party, 
incorporating identity strength measures 
will likely improve targeting for minorities within a party 
(young, Black, Hispanic republicans / older, white democrats)
but comes with the cost of additional questions. 

If the goal is to understand why demographics fail to better predict party id, 
we think that identity strength measures can contribute to the conversation. 
They demonstrate that minorities within a party may be internally de-emphasizing 
their connections to the groups commonly associated with their chosen out-party.
Future work should explore this more fully by obtaining data on the perceptions 
of which groups are best represented by each party. 

Here, our goal was to thoroughly understand party identification 
in terms of the demographic groups that one belongs to 
and how those groups are incorporated into identity through various
measures of identity strength. 
In order to form a comprehensive idea of identity, we designed a
complex survey which required approximately 30 minutes to complete. 
This design makes sense for a research setting but is likely cumbersome 
for field work. 
We explored how to reduce the survey length for such a setting, 
while minimizing the loss in predictive performance. 
We propose that incorporating Group Association, 
Group Fate, and Group Support for race, age, and gender 
into surveys will likely maximize the predictive benefits with minimal 
additional costs. 

There has been concern that the measure of Group Fate, 
which was designed for Black respondents, 
would apply to other groups as well \citep{sanchez2016taking}. 
Like others, we find that other groups experience 
perceptions of shared fate \citep{gershon2019intersectional,bejarano2021shared};
we find that 
Group Fate is a predictive factor across groups. 
While there are drawbacks to a predictive framework, 
these findings point to future work exploring measures of identity.
For example, to the best of our knowledge we are the first to propose Group Support, 
a measure which we found to be highly predictive of party id.
We propose that this measure be tested thoroughly to determine if 
it applies in other situations of understanding politics, groups, 
and identity.

\bibliographystyle{tfcad}
\bibliography{interactcadsample}

\begin{table*}[htp]
\centering
\begin{minipage}[t]{.49\textwidth}
\begin{tabular}{l  b| b | a }
\hline
\mc{1}{}  & \mc{1}{\thead{Predicted \\ Gop}} & \mc{1}{\thead{Predicted \\ Dem}} & \mc{1}{\thead{Predicted  \\Ind}} \\
\hline
\rowcolor{white}
Actual Gop & \cellcolor{blue!25}1861 & 1074 & 15  \\ \hline
\rowcolor{white}
Actual Dem  & 1230 & \cellcolor{blue!25}2103 & 15  \\ \hline
\rowcolor{white}
Actual Ind & 402 & 379 & \cellcolor{blue!25}16  \\ \hline
\rowcolor{white}
F1 Score & 0.58 & 0.61 &
\end{tabular}
\end{minipage}

\caption[Confusion matrix for ANES predictions]{The  confusion matrix in predicting party identification from ANES with a standard model and demographics. }
\label{fig:ANES_Prodege}
\end{table*}
\clearpage
\clearpage

\clearpage

\clearpage
\clearpage
\clearpage

\listoffigures

\end{document}